\documentstyle[aps,prb,multicol,epsf,graphicx]{revtex}

\newcommand \bea {\begin{eqnarray}}
\newcommand \eea {\end{eqnarray}}
\newcommand \be {\begin{equation}}
\newcommand \ee {\end{equation}}
\newcommand \eps {\epsilon}
\newcommand \bi {\bibitem}
\newcommand{\eq}[1]{~(\ref{#1})}
\newcommand \s {\sigma}

\begin{document}

\title{Statistics of lowest droplets in two-dimensional Gaussian
Ising spin glasses} 

\author{M. Picco$^{(1)}$, F. Ritort$^{(1,2)}$ and M. Sales$^{(2)}$} 

\address{(1) LPTHE, Universit\'e Pierre et Marie Curie, Paris VI
        et Universit\'e Denis Diderot, Paris VII\\
        Boite 126, Tour 16, 1$^{\it er}$ \'etage, 4 place Jussieu,
        F-75252 Paris Cedex 05, France\\
(2) Departament de F\'{\i}sica Fonamental,
 Facultat de F\'{\i}sica, Universitat de Barcelona\\ Diagonal 647,
 08028 Barcelona,Spain
}

\maketitle

\begin{abstract}
A new approach to determine the value of the zero-temperature thermal
exponent $\theta$ in spin glasses is presented. It consists in
describing the energy level spectrum in spin glasses only in terms of
the properties of the lowest energy droplets and the {\em lowest
droplet exponents} (LDEs) $\lambda_l,\theta_l$ that describe the
statistics of their sizes and gaps. We show how these LDEs yield the
standard thermal exponent of droplet theory $\theta$ through the
relation, $\theta=\theta_l+d\lambda_l$. The present approach provides
a new way to measure the thermal exponent $\theta$ without any
assumption about the correct procedure to generate typical low-lying
excitations as is commonly done in many perturbation methods including
domain wall calculations. To illustrate the usefulness of the method
we present a detailed investigation of the properties of the lowest
energy droplets in two-dimensional Gaussian Ising spin glasses.  By
independent measurements of both LDEs and an aspect-ratio analysis, we
find $\theta(2d)\simeq -0.46(1)<\theta_{DW}(2d)\simeq -0.287$ where
$\theta_{DW}$ is the thermal exponent obtained in domain-wall
theory. We also discuss the origin of finite-volume corrections in the
behavior of the LDE $\theta_l$ and relate them to the finite-volume
corrections in the statistics of extreme values. Finally, we analyze
some geometrical properties of the lowest energy droplets finding
results in agreement with those recently reported by Kawashima and
Aoki~\cite{KA}. All in all, we show that typical large-scale droplets
are not probed by most of the present perturbation methods as they probably do
not have a compact structure as has been recently suggested. We
speculate that a multi-fractal scenario could be at the roots of the
reported discrepancies on the value of the thermal exponent $\theta$ in 
the two-dimensional Gaussian Ising spin glass.
\end{abstract}


\section{Introduction.}
\label{intro}

Despite three decades of work in the field of spin glasses major
issues related to their low-temperature behavior still remain
unresolved \cite{REVIEW}. Although important achievements have been
obtained in the understanding of mean-field theory \cite{MFT} the
appropriate treatment beyond mean-field to include short-range
interactions is yet to be found. The absence of a successful
analytical approach to deal with this problem corroborates the present
state of our knowledge, often misguided by a non-accurate, if not
confusing, interpretation of the numerical data. This situation has
generated a hot debate about the correct physical interpretation of
the available numerical data.  Leaving aside the long-standing
controversy whether replica symmetry breaking is or not a good
description of the spin-glass phase \cite{RSB}, there are still
unresolved issues which are not as striking but evidence our ignorance
about some fundamental questions.

One among these problems is the correct value of the thermal exponent in
two-dimensional (2d) Gaussian Ising spin glasses (GISG). This question
has received attention from time to time during the last two decades but
not enough to settle it definitively and explain the origin of some of
the reported discrepancies. The study of the low-$T$ properties of the
2d GISG starts with the work by McMillan who proposed \cite{MILLAN1}
that thermal properties in spin glasses are determined by the scaling
behavior of the typical largest excitations (commonly referred as
droplets) present in the system.  This idea has been further elaborated
and extended to deal with equilibrium and dynamical properties of spin
glasses in a scenario nowadays referred to as droplet model
\cite{DROPLET}. The low-$T$ behavior in spin glasses is determined by a
spectrum of large scale gapless droplets with typical length $L$ and
energy cost $E\sim L^{\theta}$, $\theta$ being the thermal exponent. As
these droplets correspond to flipping some domains of spins (assumed to
be compact clusters), the energy cost of these excitations arises from
the set of unsatisfied bonds on their surface. The striking low-$T$
behavior in spin glasses arises from multiple energy cancellations
occurring at the surface of the droplet.  These cancellations can be
seen as the result of a competition between energy and entropy effects:
as the droplet becomes progressively larger there are more available
conformations for the surface to minimize the energy cost of the
unsatisfied bonds. In the absence of cancellations one would expect
$\theta=(d-1)/2$. However, as these cancellations are very important,
the inequality $\theta<(d-1)/2$ holds and $\theta$ is by far less than
the maximum value $(d-1)/2$.  The value of the thermal exponent $\theta$
characterizes the low-$T$ critical behavior as it is related to the
correlation length exponent $\nu$ where $\xi\sim T^{-\nu}$ by the
identity $\nu=-1/\theta$. McMillan also used domain-wall renormalization
group ideas to introduce a practical way to determine the leading energy
cost of these low-lying large-scale excitations \cite{MILLAN2}. The
method consists in measuring the energy defect of a domain-wall spanning
the whole system obtained by computing the change of the ground state
energy when switching from periodic to anti periodic boundary conditions
in one direction.  Several works have used McMillan's method to
determine the value of $\theta$ in two and three dimensions
\cite{VARIOS,RIEGER}. Hereafter, in order to keep the discussion as
clear as possible, we will denote by $\theta_{DW}$ the estimate of the
exponent $\theta$ obtained by domain-wall calculations.  The initial
value for $\theta_{DW}$ reported by McMillan is $\theta_{DW}=-0.281(5)$
for pretty modest lattice sizes $L=3-8$. Recent numerical results with
much more powerful algorithms have reached sizes $L\simeq 500$ and
confirmed the initial result with much larger
accuracy~\cite{HY01,CBM02,A02} $\theta_{DW}=-0.287(4)$.  These studies
would definitively close the problem if it were not by the existence of
other alternative estimates of the exponent $\theta$, largely consistent
among them, which yield a quite different value $\theta\simeq
-0.47(2)$. We will denote this estimate by $\theta_{TF}$ as several of
these methods use transfer matrix~\cite{KHS}. However, a word
of caution is necessary here as the Monte Carlo method and other
approaches that are not based on transfer matrix methods report values
compatible with that estimate. For instance, Kawashima and Aoki used
another method to estimate the stiffness exponent~\cite{KA}. The idea is
to generate a droplet inside a box of size $L\times L$ that includes a
fixed central spin, with the following procedure. First, the ground state
is found with a standard algorithms (we will denote it by the reference
configuration). Afterwards, the spins at the boundaries of the box are
fixed and the central spin is forced to flip respect to the reference
configuration. The droplet of minimum energy that includes the central
reversed spin and does not touch the boundaries is computed. The
spanning length of the droplets generated in this way allows to define
the fractal dimension of both the surface (or perimeter for the two
dimensional case) and the volume. It is found that these minimum energy
droplets have a fractal volume dimension smaller than 2 and the thermal
exponent is $\theta=-0.42(5)$ in agreement with results obtained from MC
methods~\cite{MC} and heuristic optimization algorithms~\cite{JAP}. A
similar study of minimum energy clusters in the three-dimensional
Edwards-Anderson model also reports evidence that $\theta_{DW}$ is an
upper bound to the actual value of the thermal exponent~\cite{LBMM}.

The accuracy of previous estimates is poorer than the values obtained
through the domain-wall method as they deal, in one way or another, with
all possible excitations and not only with the calculation of ground
state energies. More recently, another method has been used to estimate
the value of $\theta$.  It consists in perturbing the original
Hamiltonian ${\cal H}_0$ with a term $\eps {\cal P}$, where ${\cal P}$
stands for the perturbation and $\eps$ for its intensity. For example,
$P$ can be the overlap between the actual configuration and the ground
state of the original Hamiltonian ${\cal H}_0$. As $\eps$ varies the new
ground state of the total Hamiltonian ${\cal H}={\cal H}_0+\eps {\cal
P}$ remains unchanged until a certain value $\eps=\eps_c$ is reached
where a excited energy level of ${\cal H}_0$ becomes the new ground
state of ${\cal H}$. The overlap between the old and the new ground
states as well as the value of the shifting energy provoked by the
perturbation links its energy cost $E$ with its size providing another
way to estimate $\theta$. We will denote by $\theta_P$ the estimate
obtained in this way.  This method has been recently used in the 2d GISG
by Hartmann and Young~\cite{HY02} reporting the following value
$\theta_P\approx -0.31$. Although slightly more negative than
$\theta_{DW}$, both $\theta_{DW},\theta_P$ appear to be statistically
compatible. Yet more accurate estimates are needed to confirm
whether $\theta_P=\theta_{DW}$.

This last method and the domain-wall method have in common the same
feature, i.e. they perturb the original Hamiltonian in one way or
another to probe the characteristic energy of excitations that are
supposed to be the typical ones that determine the low-$T$ thermodynamic
properties. In fact, the estimate $\theta_{DW}$ can be considered as a
particular example of $\theta_P$, where the perturbation consists in
reversing all the bonds in one of the surfaces of the box.  This raises
the important question whether the different estimates of $\theta_P$,
obtained by considering different class of perturbations, are
different. The question is rather subtle as there are numerical
indications that indeed this could be the case. For
instance~\cite{RIEGER}, measurements of $\theta_P$ where the
perturbation is a uniform magnetic field yield a value
$\theta_P=-0.48(1)$ compatible with the other competing set of values
$\theta_{TF}$.

How is that the value of the exponent $\theta_P$ could depend on the
type of perturbation? This is a very difficult question to answer as
our present knowledge is inadequate. We can offer only speculative
answers. Strong discrepancies among different types of perturbations
could arise if a multi-fractal scenario governs the statistics of
excitations in spin glasses. By definition, in all perturbation
methods the probed large scale droplets are those which minimize the
energy cost but constrained to maximize the value of the perturbation
for the selected droplets. Therefore, among all possible large-scale
low-lying droplets the perturbation method selectively probes those
that maximally overlap with the perturbation. A dependence of the
value $\theta_P$ on a given class of perturbations could arise if the
perturbation selectively probes one or another topological property
of the droplet. This rather awkward multi-fractal scenario is not
new in the field of disordered systems. Multifractality is known to be
present in the localization problem in the strongly disordered regime. A
multi-fractal scenario would imply the existence of different critical
exponents at $T=0$ depending on the type of perturbation applied.  On
the other hand, the fact that the value estimated for $\theta_P$ when
the perturbation is a uniform magnetic field appears to be consistent
with the value $\theta_{TF}$, suggests that maybe some types of
perturbation can probe the relevant excitations while others may
not. These good observables, which probe the typical excitations, could
be called neutral observables in the same spirit as this term has been
coined to describe observable dependences of the fluctuation-dissipation
ratio (i.e. the effective temperature) in glassy systems. Concomitant,
this ``perturbation class dependence'' issue is presently also debated in
the different (but related to a certain degree) field of glassy
dynamics.

If the hypothetical multi-fractal scenario holds, then we must
face the question about what is the correct procedure to determine
the thermal exponent $\theta$. As $\theta$ determines the free
energy cost of droplets, the natural answer is that $\theta$ is given by
the lowest value among all possible estimates,
\be
\theta={\rm min}_{\cal P}\lbrace{\theta_{\cal P}\rbrace} \; .
\label{eqP}
\ee
With the present available data this relation suggests that the estimate
$\theta_{TF}$ is the correct value of the thermal exponent and that
$\theta_{DW}$ as well as many other estimates $\theta_P$ are only 
upper bounds to the true value.

The question we want to address in this paper is the following. Is it
possible to devise a method that is alternative to current perturbation
methods, in which excitations are not selectively probed by the
perturbation, but selected only according to the correct balance
between energy and entropy? The main purpose of this paper is to show
that the analysis of the statistics of the first or lowest excitations
gives a positive answer to this question.  As we will see, the method
we propose in this paper yields a consistent estimate of $\theta$
compatible with the value $\theta_{TF}$, therefore supports the result
that $\theta_{DW}$ and many other $\theta_P$ are only upper bounds to
the actual value of $\theta$.  A preliminary account of these results
has already appeared in \cite{NEXT}.

The paper is divided as follows. Sec.~\ref{basis} describes the basis
of the lowest droplet approach and introduces the lowest droplet
exponents. Sec.~\ref{2d} shows the results obtained in the 2d GISG.
Sec.~\ref{best} analyzes a method to extract the value of the thermal
exponent $\theta$.  Sec.~\ref{ARA} presents a more powerful method to
extract the value of the lowest droplet exponents based on an
aspect-ratio analysis. Sec.~\ref{fse} discusses the origin of the
finite-volume corrections to the value of the lowest droplet exponent
$\theta_l$ as a problem of corrections in the statistics of extreme
values.  Sec.~\ref{compact} analyzes some topological properties of
the lowest droplets. Finally Sec.~\ref{conclusions} presents the
conclusions. There are also two technical appendixes:
Appendix~\ref{upperbound} presents the heuristic argument that
$\theta_l=-d$ for Gaussian spin glasses, and
Appendix~\ref{transfermatrix} explains the transfer matrix method we
used to obtain the lowest droplets.

\section{Basis of the lowest droplet approach.}
\label{basis}

The purpose of this work is to show an alternative approach to
determine the low-$T$ behavior of spin glasses by studying the size
and energy spectrum of the lowest excitations by introducing two
exponents ($\lambda_l$ and $\theta_l$) needed to fully characterize
the zero-temperature fixed point. All along the paper we will denote
this exponents as {\em lowest energy droplet} exponents, or {\em
lowest droplet} exponents in short, and that we will abbreviate as LDEs. The
exponent $\lambda_l$ is the most important one and describes the
probability to find a large-scale lowest excitation spanning the whole
system, while the exponent $\theta_l$ describes the system-size
dependence of the average energy cost of these lowest excitations.

The underlying theoretical background of the approach is the
following. To investigate the leading low-temperature behavior in spin
glasses let us consider expectation values for moments of the order
parameter by keeping only the ground state and the first or lowest
excitation. This approach was introduced in \cite{RS} and can be shown
to capture the low-temperature behavior at the leading order.  The
method that investigates the low-$T$ properties based on a restricted
analysis of the spectrum to the absolute lowest excitations has been
also used for the study of the localized phase in the disordered
Anderson model~\cite{GEFEN}. The present paper can be seen as the
applicability of these ideas to the spin-glass case. At the end of the
paper (see Sec.~\ref{conclusions}) we will give reasons supporting the
validity of our approach.

To generate the spectrum of lowest excitations we consider the
following procedure. Let us consider a set of ${\cal N}_s$ samples and
for each of them we determine both the configurations of the ground
state and the lowest excitation. For a spin model the lowest
excitation has $v$ spins overturned with respect to the ground state
(so the overlap between the ground and that excited state is
$q=1-2v/V$, $V$ being the volume of the system) and with energy cost
or gap $E$. It can be easily proved that the lowest excitation must be
a connected cluster which we will generically call the lowest
droplet. If $v_s$ and $E(s)$ denote the volume and excitation energy
of the lowest droplet for sample $s$, in the limit where ${\cal N}_s$
is sent to infinity, we can define the following joint probability
distribution
\be
P(v,E)=\frac{1}{{\cal N}_s}\sum_{s=1}^{{\cal
N}_s}\delta(v-v_s)\delta(E-E(s))\; .
\label{eq2}
\ee
Using the Bayes theorem, this joint probability distribution can be
written as $P(v,E)=g_v\hat{P}_v(E)$ where
\be
\sum_{v=1}^{\frac{V}{2}}g_v=1~~~~~;~~~~\int_0^{\infty}dE\hat{P}_v(E)
=1~~~\forall v \; .
\label{eq_norm}
\ee
$g_v$ is the probability to find a
sample such that its lowest droplet has volume $v$ and $\hat{P}_v(E)$
is the conditioned probability for that droplet to have a gap equal
to $E$. In what follows, we separately discuss the scaling behavior of
both distributions $g_v,\hat{P}_v(E)$.

Before continuing, and for sake of clarity, let us make an important
digression about nomenclature. There are two {\em volumes} involved in
the problem: the volume $v$ of the lowest excitation and the volume $V$
of the lattice. If not stated otherwise we will refer to the volume $v$
as the {\bf size} of the excitation while {\bf volume} will generally
refer to the lattice volume $V$. Thus, when we speak about finite-size
excitations we usually refer to excitations with $v$ finite, and
finite-volume corrections (which we will sometimes abbreviate as FVC)
will refer to the corrections affecting the distribution \eq{eq2} due to
the finite volume $V$ of the lattice.

\subsection{The lowest droplet exponent $\lambda_l$.}
\label{lamdal}

The simplest scenario for the size distribution of the lowest droplets
is that all sizes occur with uniform probability. The normalization
condition~(\ref{eq_norm}) imposes $g_v\sim 1/V$. This situation is
encountered in the 1d GISG~\cite{DROPLET,RS} with both free and
periodic boundary conditions. However, in the most general situation,
this does not hold and low energy droplets are found with
a probability that depends on their size $v$. The simplest and most
general way to incorporate such a dependence is to assume an ansatz
solution for $g_v$ that factorizes into a power law
$A/V^{\lambda_l+1}$ with $\lambda_l>0$ and a coefficient $A\equiv
G(q)$ which depends only on the overlap $q$ between the ground state
and the lowest droplet, \be g_v=\frac{G(q)}{V^{\lambda_l+1}} \; .
\label{eq4}
\ee
The behavior of $G(q)$ can be guessed in both limits $q\to 1$ (the
case $q\to -1$ is equivalent in models with time-reversal symmetry which
are those we are considering here) and $q\to 0$,
\begin{eqnarray}
G(q\to 0)\to {\rm constant}\label{q0}
\\
G(q\to 1)\to \frac{1}{(1-q)^{\lambda_l+1}}\label{q1} \; .
\end{eqnarray}
The first relation describes the scaling behavior for the number of
droplets whose size scales with the total volume of the system. As these
can only depend on the volume $V$, $G(0)$ must converge to a
constant. The second relation is consequence of the fact that the number
of droplets with finite size $v$ cannot depend on $V$ in the large $V$
limit as these are not affected by the boundaries. On the other hand,
the distribution of finite size droplets $g_v$ is self-similar as can be
seen by inserting (\ref{q1}) in (\ref{eq4}) and using the relation
$q=1-2v/V$. This yields $g_v\sim 1/v^{\lambda_l+1}$, the same relation as for
the large scale limit (\ref{q0}) where $g_V\sim 1/V^{\lambda_l+1}$. A
simple expression that interpolates both limits is given by,
\be
G(q)=\bigl(A+\frac{B}{(1-q)^{\lambda_l+1}}\bigr)\; .   
\label{eq_inter}
\ee 
Note however that, despite its simplicity, expression (\ref{eq_inter})
is only an interpolation and the most we can say about $G(q)$ concerns
its asymptotic behaviors (\ref{q0}-\ref{q1}).

The ansatz (\ref{eq4}), applied only to large-scale
excitations, was proposed in \cite{RS}. Note that although $g_v$ is
defined for discrete volumes, in the limit $V\gg 1$, the values of $q$
for consecutive droplet sizes $v\to v+1$ become equally spaced by
$\Delta q=2/V$. Therefore, in the limit, $V\gg 1$, the function
$g(q)=\frac{V}{2}g_v$ becomes a continuous function if expressed in terms
of the variable $q$ instead of the integer variable $v$,
\be
g(q)=\frac{1}{2V^{\lambda_l}}G(q)\; .   
\label{eq5}
\ee 
A word of caution is in order. Although (\ref{eq4}) diverges for
$q=1$, leading apparently to a violation of the normalization condition
(\ref{eq_norm}) for $g_v$, it must be emphasized that no excitation
has $q=1$ so there is a maximum cutoff value $q^*=1-2/V$ corresponding
to one-spin excitations. For instance, if we insert (\ref{eq5}) into the
normalization condition for $g(q)$ we get in the large $V$ limit,

\be
\int_0^{q^*=1-2/V}g(q)dq=1\rightarrow
\frac{A-B/\lambda_l}{2V^{\lambda_l}}+\frac{B}{2^{\lambda_l+1}\lambda_l}=1 \; ,
\label{eq6}
\ee
%
implying $\lambda_l\ge 0$ as expected since otherwise the
normalization would not be possible in the large-$V$ limit. The
divergent term ($q\to 1$) in (\ref{eq5}) shows that for $\lambda_l>0$
one-spin excitations are the most large in number among the whole
spectrum of sizes.  In fact, from (\ref{eq4}), $g(1)\simeq O(1) \gg
g(V/2)\simeq 1/V^{\lambda+1}$, so the majority of excitations have a
finite size. However the average excitation size
\be
\overline v=\sum_{v=1}^V v g_v\to_{V\to\infty} V^{1-\lambda_l} \; ,
\label{eqv}
\ee
diverges in the $V\to \infty$ limit and differs from the typical
excitation volume $v_{\rm typ}\sim O(1)$. Relation (\ref{eqv}) provides
a way to measure the exponent $\lambda_l$ alternative to the
use of the scaling behavior (\ref{eq4}).

\subsection{The lowest droplet exponent $\theta_l$}
\label{thetal}

The analysis of the gap distribution $\hat{P}_v(E)$ goes along the
same lines as we did for the distribution $g_v$, but with one
important difference. As the gap $E$ describes the lowest among all
possible excitation energies, it has to scale in the same way for all
droplet sizes {\em independently} on their size (and, in particular,
whether these are finite-size or large scale droplets). This statement
refers to a scenario which hereafter we will call ``the random
energy-size droplet scenario'' (RESD scenario) to specifically
indicate that the distribution of the lowest energies of droplets is
independent of their size. Mathematically it can be expressed as,
\be
\hat{P}_v(E)=\hat{P}(E)~~~~,~~~~\forall v \; .
\label{REM}
\ee
In addition, we follow the standard droplet model and assume that the
spectrum is gapless and defined by an exponent $\theta_l$ which
describes the characteristic energy of the lowest droplets whatever
their size or overlap $q$ with the ground state.  If the scaling
function $\hat{P}_v(E)$ is independent of $v$ it follows immediately
that the non-conditioned or size-averaged gap probability distribution
\be
P(E)=\sum_{v\ge 1} g_v\hat{P}_v(E)=\sum_{v\ge 1}
g_v\hat{P}(E)=\hat{P}(E)
\label{eqpe}
\ee
where we used (\ref{REM}) and the normalization condition
(\ref{eq_norm}) for $g_v$. From now on, if not stated otherwise,
we will always refer to the size-averaged probability distribution
$P(E)$ with the clear understanding that it coincides with any of the
conditioned distributions $\hat{P}_v(E)$. As the spectrum of lowest
excitations is gapless, the normalized distribution $P(E)$ has the
following scaling behavior,
\be
P(E)=\frac{1}{L^{\theta_l}} {\cal P}\Bigl (\frac{E}{L^{\theta_l}}\Bigr)\; . 
\label{eq7}
\ee
We stress that the exponent $\theta_l$ is completely different
from the standard thermal exponent (see next section) as they
describe totally different excitations. The thermal exponent $\theta$
describes the energy-length relation for droplets typically excited at
finite temperatures while the lowest energy exponent $\theta_l$
describes the droplets that are separated by the smallest gap respectively to
the ground state, so that, in general, $\theta_l\le\theta$.

We will argue below in Sec.~\ref{theta} that $\theta_l=-d$ for a generic
class of spin-glass systems with coupling distributions with finite
weight at zero gap. In addition, this relation will provide an
alternative interpretation of the lower critical dimension in terms of
the exponent $\lambda_l$ introduced in Sec.~\ref{lamdal} describing
the properties of the spectrum of sizes of the lowest droplets.

\subsection{The standard thermal exponent $\theta$}
\label{theta}

Now we want to show how the exponents $\lambda_l$ and $\theta_l$ combine
to give the usual scaling exponent $\theta$ describing the energy cost
of {\em typical} thermal excitations in droplet theory.
There are several ways to show this result.  For simplicity,
here we exemplify this relation by analyzing the low-$T$ behavior of the
second moment of the spin-glass order parameter
at the
order linear in $T$ by keeping only the first excitation. If
$q_{\lbrace\sigma,\tau\rbrace}=\frac{1}{V}\sum_i\sigma_i\tau_i$ denotes the overlap between two replicas
(i.e. configurations of different systems with the same realization of
quenched disorder), then the expectation value $\overline{\langle
q^2\rangle}$ can be written as follows \cite{RS},

\be
\overline{<q^2>}=1-\frac{2}{V^2}\sum_v\int_0^{\infty}dEP(v,E)
v(V-v){\rm sech}^2 \left(\frac{E}{2T}\right),
\label{eq1}
\ee

\noindent
where $P(v,E)$ is given by (\ref{eq2}). A low-temperature expansion of
(\ref{eq1}) \cite{RS,NEXT} up to linear order in $T$ yields,

\be
\overline{<q^2>}=1-\frac{4T}{V^2}\sum_{v=1}^V\,g_v\hat{P}_v(0)v(V-v)
\label{eq3}
\ee

\noindent
which shows that the leading behavior is determined by both $g_v$ and
the density of states at zero gap $\hat{P}_v(0)$. In the standard droplet model,
it is generally assumed that typical low energy droplets have an average
size $\overline v=\sum_v v g_v\sim V$ of the order of the system size
(such as those generated by DW perturbation) and finite
weight at zero gap $\hat{P}_V(0)\sim 1/L^{\theta}$ where $\theta$ is
the thermal exponent. In principle, a single exponent $\theta$
describes the scaling behavior of typical large-scale droplets with volume
$v\propto V$ and determines the zero-temperature
critical behavior. As these large-scale droplets are typical they occur with
finite (therefore independent of $V$) probability $g_V\sim O(1)$ while
small scale droplets are simply irrelevant $g_{v\sim O(1)}\sim 0$. 
This yields,
\be
\overline{<q^2>}=1-c\frac{T}{L^{\theta}},
\label{eqq2_1}
\ee
where $c$ is a non-universal stiffness constant related to the
particular model.  One of the most relevant results from the ansatz
(\ref{eq4}) is that both small and large scale excitations contribute
to low-temperature properties. In general, let us consider any
expression (such as (\ref{eq3})) involving a sum over all possible
volume excitations. Restricting the sum to the large-scale droplets
($v/V$ finite) the net contribution to such sum is proportional to
$Vg_V\hat{P}_{V}(0)\propto L^{-\theta_l-d\lambda_l}{\cal P}(0)$ (where
${\cal P}$ is the scaling function appearing in (\ref{eq7})).  Coming
back to (\ref{eq3}) and using (\ref{eq4}) and (\ref{eq7}), we note that
both small and large-scale excitations yield a contribution to (\ref{eq3}) of the
same order and given by,
\be
\overline{<q^2>}=1-c_l\frac{T}{L^{\theta_l+d\lambda_l}} \; ,
\label{eqq2_2}
\ee
where $c_l$ is another constant (different from the constant $c$
appearing in (\ref{eqq2_1})). Identifying both relations (\ref{eqq2_1}) and
(\ref{eqq2_2}) we obtain the general relation,
\be
\theta=\theta_l+d\lambda_l \; .
\label{general}
\ee
This relation shows how the value of $\theta$ can be computed from
$\lambda_l$ and $\theta_l$. Through the study of a specific example, we
will see later that the exponents $\theta_l$ and $\lambda_l$ have strong
finite-volume corrections arising from the corrections present in 
the statistics of the extreme values. However, we will present
alternative routes to overcome this dependence and provide an accurate 
estimate of $\theta$.

Now we come back to the aforementioned argument at the end of
Sec.~\ref{thetal} claiming that in the large-volume limit $\theta_l$
must converge to the value $-d$ in the case of coupling distributions
with finite gap at zero coupling.  The details of the argument are
shown in appendix~\ref{upperbound}. The argument has two parts. First,
it is proved that one-spin excitations provide an upper bound for the
LDE $\theta_l$. Then it is argued that this upper bound holds also for
any finite-size excitations (such as two-spin clusters, three-spin
clusters, and so on). We will see below how this result is supported
by the numerical analysis of the data.  Let us also note that this
result, in a RESD scenario (see Sec.~\ref{thetal}) can be linked to
the linear dependence of the specific at low temperatures, a result
widely accepted, but that has been revisited recently in~\cite{NEXT2} to
show that it has strong FVC due to the systematic FVC present in the
value of $\theta_l$. Inserting $\theta_l=-d$, (\ref{general}) becomes,
\be
\theta=d(\lambda_l-1) \; .
\label{general2}
\ee
This relation provides a way to distinguish the lower critical dimension
$d_{lcd}$ in terms of the average size distribution of the lowest
droplets. According to (\ref{eqv}) the relation $\lambda_l(d_{lcd})=1$
distinguishes a regime where the average size of the lowest droplet
grows with the volume of the system to a regime where the
average size of the lowest droplet is finite, 
\bea
d<d_{lcd}: \lim_{V\to\infty} \overline{v}(V)=
\infty~~;~~\lambda_l<1,\theta<0\label{d<}\\
d>d_{lcd}: \lim_{V\to\infty} \overline{v}(V)= 
O(1)~~;~~\lambda_l>1,\theta>0 \; .
\label{d>}
\eea
The marginal case $\lambda_l=1,\theta=0$ is specially interesting as
the average size $\overline{v}$ could be finite or diverge with the
size but slower than a power law. This scenario corresponds to the
mean-field behavior as replica-symmetry is broken in both the standard
RSB~\cite{RSB} or in the TNT~\cite{TNT} (standing for trivial-non
trivial) scenarios. Therefore, the study of the size spectrum of the
lowest excitations in spin glasses can be very useful to find out the
correct value of the thermal exponent in models without a finite-$T$
transition (such as the 2d GISG) as well as establishing the correct
low-$T$ scenario in models with a finite-$T$ transition.  In the next
section we apply all these ideas to evaluate the thermal exponent for
the $2d$ GISG.

\section{Statistics of the lowest energy droplets in the 2d GISG}
\label{2d}

Several numerical works have recently searched for low-lying
excitations in spin glasses using heuristic algorithms
\cite{MARTIN}. But, to our knowledge, no study has ever presented
exact results about the statistics of lowest excitations. We have
exactly computed ground states and lowest excitations in
two-dimensional Gaussian spin glasses defined by

\be
{\cal H}=-\sum_{i<j}J_{ij}\,\s_i\,\s_j \; ,
\label{eq0}
\ee

\noindent
where the $\s_i$ are the spins ($\pm 1)$ and the $J_{ij}$ are quenched
random variables extracted from a Gaussian distribution of zero mean
and unit variance. These have been computed by using a transfer
matrix method working in the spin basis. Representing each spins state
by a weight and a graduation in the energy we can build explicitly the
ground state by keeping the largest energy and, by subsequent
iteration, the first excitation and so on (see
Appendix~\ref{transfermatrix} for the details on how we compute these
quantities). The continuous values for the couplings assures that
there is no accidental degeneracy in the system (apart from the
trivial time-reversal symmetry $\sigma\to -\sigma$). Calculations
have been done in systems with free boundary conditions in both
directions (FF), periodic boundary conditions in both directions (PP)
and free boundary conditions in one direction but periodic in the
other (FP). In all cases we find the same qualitative and
quantitative results indicating that we are seeing the correct
critical behavior.

We have found ground states and lowest droplets for systems ranging
from $L=4$ up to $L=11$ for PP and up to $L=16$ for FP and FF. The
number of samples is very large, typically $10^6$ for all sizes. The
large number of samples assures us that many samples have large-scale
droplets as first excitations. This provides us with good statistics
to properly analyze the sector of large-scale excitations. The large
number of samples requires a big amount of computational time so that
calculations were done in a PC cluster during several months. For each
sample we have evaluated the volume of the excitation $v$ (and hence
the overlap $q=1-2v/V$ between the ground state and the first
excitation) and the gap $E$. From these quantities we can construct
the $g_v$ and the $\hat{P}_v(E)$.

\begin{figure}[tbp]
\begin{center}
\includegraphics*[height=8.5cm]{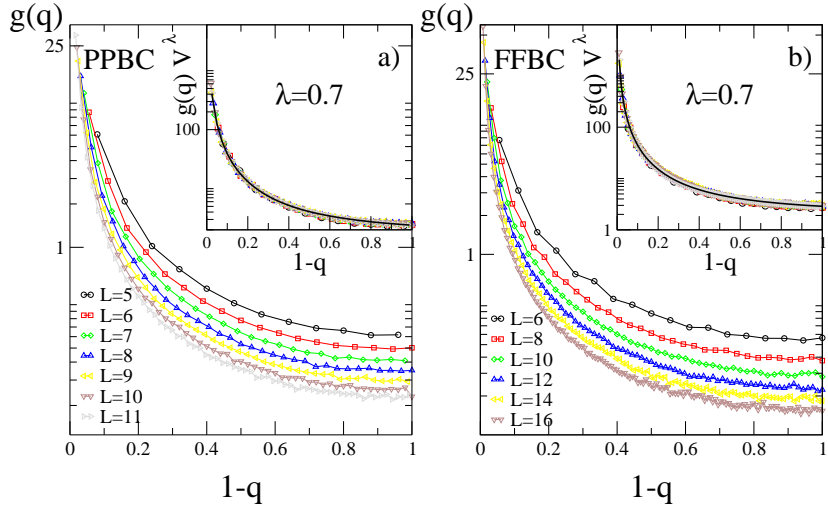}
\vskip -0.3in
\caption{$g(q)$ versus $1-q$ for the PP (left panel) and FF case
(right panel) for different lattice sizes $L=5-11$ (PP) and L=6-16
(FF) from top to bottom.  In both insets we plot the scaling function
$g(q) V^{\lambda}$ versus $1-q$ with $\lambda=0.7$.
\label{FIG1}}
\end{center}
\end{figure}

In Fig.~\ref{FIG1} we show $g(q)=\frac{V}{2}g_v$ as function of $q$
for different sizes in the PP and FF cases. We can clearly see that
there are excitations of all possible sizes but, as discussed in the
paragraph following (\ref{eq6}), the typical ones which dominate by
far are single spin excitations. To have a rough idea of the number of
rare samples giving large scale excitations let us say that nearly
half of the total number of samples have one-spin lowest excitations,
whereas less than $10\%$ of the samples have lowest excitations with
overlap $q$ in the range $0-0.5$. This disparity increases
systematically with size. For the lattice sizes explored the typical
number of large-scale droplets is in the range $10^4-10^5$ which is,
indeed, quite good to have a good sampling of the sector corresponding
to large scale excitations.  A detailed analysis of the shape of $g_v$
reveals that it has a flat tail for large-scale excitations and a
power-law divergence for finite-size excitations.  The $g_v$ can be
excellently fitted by the interpolating formula (\ref{eq_inter},\ref{eq5})),
\be
g(q)=\frac{2}{V^{\lambda_l}}\left(A+\frac{B}{(1-q)^{\lambda_l+1}}\right)\; .
\label{fit}
\ee
As shown in the insets of Fig.~\ref{FIG1} a good collapse of the
scaling function is obtained with the effective exponent
$\lambda_l^{\rm eff}\simeq 0.7$ for both PP and FF cases. We also plot
the line resulting from the fit of~(\ref{fit}) with numerical data
with the following values for $A$ and $B$: PPBC: $A=1.55(3)$ and
$B=0.777(3)$; FFBC: $A=2.02(3)$ and $B=0.85(1)$. Note that the fit is
excellent and is hardly distinguishable from the points. The value of
$\lambda_l$ is compatible with the one obtained by fitting the average
size with the expression (\ref{eqv}) with the addition of a constant
term to account for the small-$V$ behavior, $\overline
v=C_1+C_2V^{1-\lambda}$. The same exponent $\lambda_l$ can be
estimated by measuring the ratio $g(V/2)/g(1)\sim
D_1+D_2V^{-1-\lambda}$. In both cases we get an effective exponent
$\lambda_l^{\rm eff}=0.70(5)$ as best fitting value.

\begin{figure}[tbp]
\begin{center}
\includegraphics*[height=0.4\columnwidth]{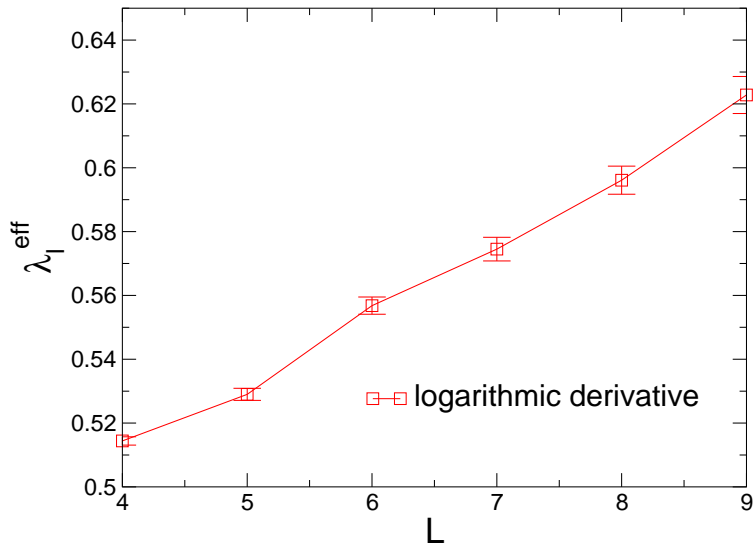}
\caption{Effective lowest droplet exponent $\lambda^{\rm eff}_l$  versus $L$ 
for the PP case, computed using logarithmic derivatives.
\label{FIG_lambda}}
\end{center}
\end{figure}

However these different estimates of $\lambda_l$
are strongly affected by finite-volume corrections (FVC). To
evidence them we have estimated an effective $L$
dependent $\lambda_l^{\rm eff}(L)$ exponent by relating the average
excitation size at consecutive sizes and using relation (\ref{eqv}),
\be \lambda_l^{\rm
eff}(L)=1-\frac{1}{d}\frac{\log\Bigl(\frac{\overline{v}(L+1)}{\overline{v}(L)}
\Bigr)}{\log\Bigl(\frac{L+1}{L}\Bigr)} \; .
\label{lambdaeff}
\ee
In Fig.~\ref{FIG_lambda} we show $\lambda_l^{\rm
eff}(L)$ in the range $L=4-11$ for the PP case. As we can appreciate
there is a systematic increase of the effective exponent as we go to
large volume sizes without any tendency to saturate. This proves that
FVC in our measurements are still big and the estimate
$\lambda_l^{\rm eff}$ used to collapse the data in Fig.~\ref{FIG1} is
still far from the asymptotic exact value.

\begin{figure}[tbp]
\begin{center}
\includegraphics*[height=7cm]{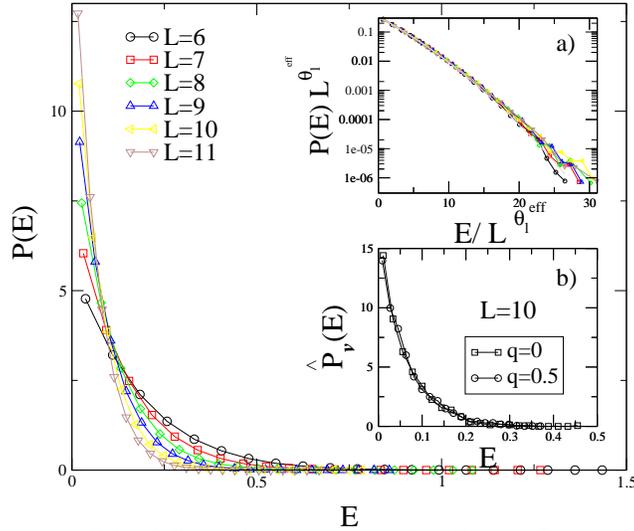}
\caption{Gap distribution $P(E)$ versus $E$ for different
lattice sizes in the PP case. In inset a) scaling obtained
from the ansatz (\ref{eq7}) with $\theta_l^{\rm eff}=-1.7(1)$. In inset b) we
show the $\hat{P}_v(E)$ for different
excitation sizes ($q=0.5, q=0$) for a lattice size $L=10$. Note 
that the distribution is independent of the size of the excitation.} 
\label{FIG2}
\end{center}
\end{figure}

After having discussed the $g_v$ we jump now to discuss the scaling
behavior of the energy gap distribution $\hat{P}_v(E)$ and its average
$P(E)$.  In Fig.~\ref{FIG2} we show $P(E)$ (main figure and inset a)
and $\hat{P}_v(E)$ (inset b) for the PP case. Similar results are
obtained for the FF and FP cases. Quite remarkably, as was already
anticipated in (\ref{REM}), the RESD scenario holds as the
distribution $\hat{P}_v(E)$ does not depend on the size $v$ of the
excitation (see inset b in Fig.~\ref{FIG2}), hence both large and
finite-size excitations are described by the same gap distribution.

In the main figure we can see how the width of distribution $P(E)$
progressively shrinks to 0 as $L$ increases. Moreover, the $P(E)$ has
an exponential shape. This is shown in the inset a) of Fig.~\ref{FIG2}
where we plot $P(E)$ in log-normal scale. Nonetheless, a detailed
examination of the tails of $P(E)$ reveals some deviations from
linearity. In Sec.~\ref{fse} we discuss the origin of these
deviations. We anticipate, though, that they are consequence of the
strong FVC in the range of sizes investigated.  In that inset we also
verify the scaling ansatz (\ref{eq7}) by showing the best data
collapse for $P(E)$ obtained with an effective exponent $\theta_l^{\rm
eff}\simeq -1.7(1)$. This is very far from the expected
value $\theta_l=-2$ discussed in the preceding Sec.~\ref{theta} and in
the Appendix~\ref{upperbound}.  A calculation of the moments of $P(E)$
(\ref{eq7}) for different values of $L$ shows that there are also
strong sub-dominant corrections to the leading scaling (\ref{eq7})
that result in corrections as large as the ones affecting the exponent
$\lambda_l$.

\begin{figure}[tbp]
\begin{center}
\includegraphics*[height=0.4\columnwidth]{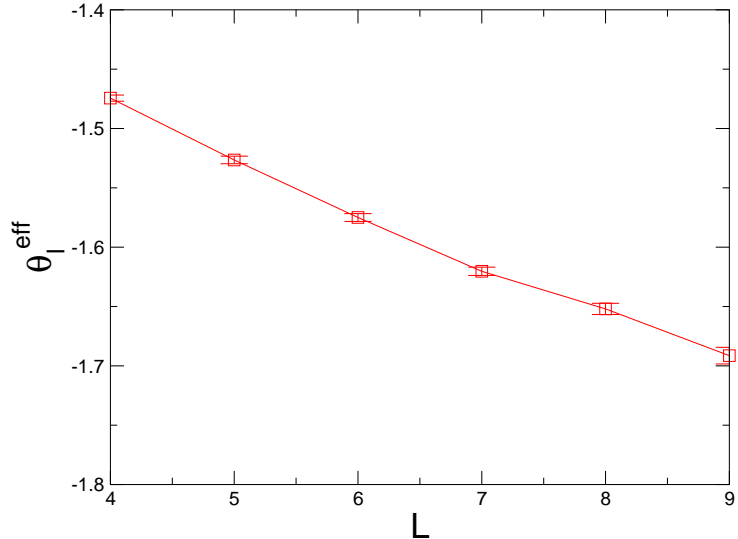}
\caption{Effective droplet exponent $\theta^{\rm eff}_l$ versus $L$ for the 
PPBC case, computed using logarithmic derivatives (see text).
\label{FIG_theta}}
\end{center}
\end{figure}

Again, to manifest the magnitude of FVC in $\theta_l$ we have
evaluated $\overline{E}(L)$, the first moment of $\hat{P}(E)$,
obtained by averaging the lowest gap over all possible droplet sizes
for different lattice sizes in the range $L=4-11$.  We have estimated
an effective $L$-dependent exponent by means of the following
expression,
\be
\theta_l^{\rm eff}(L)=\frac{\log\Bigl(\frac{\overline{E}(L 
+1)}{\overline{E}(L)}\Bigr)}{\log\Bigl(\frac{L+1}{L}\Bigr)} \; .
\label{thetaeff}
\ee
The results are shown in Fig.~\ref{FIG_theta} for the PP case. Again,
as for $\lambda^{\rm eff}_l$ (see Fig.~\ref{FIG_lambda}), we observe
that the estimated value for $\theta_l^{\rm eff}$ systematically
changes with size showing that, for the sizes we have explored, we are
still far from the asymptotic regime.

\vskip -0.15in

We can summarize the results of this section saying that both lowest
droplet exponents (LDEs) $\lambda_l$ and $\theta_l$ display strong
systematic finite-volume corrections (FVC). In principle, without
further elaboration, it is difficult to give an accurate estimate for
the thermal exponent $\theta$ using (\ref{general}). An alternative
estimate for the exponent $\theta$ could be defined from the analysis of the
fraction of large-scale excitations with $q\le 1/2$, $f(q\le
1/2)$~\cite{PETER}, which is given by
\be
f(q\le 1/2)\sim Vg(0)\sim 1/V^{\lambda_l-1} \sim
L^{d(\lambda_l-1)}\sim  1/L^{\theta} \; ,
\label{peter}
\ee
where we have used $\theta_l=-d$~\eq{general2}. Although \eq{peter}
yields estimates for $\theta$, again these are affected by strong
finite-volume corrections. In the range of sizes studied in this
paper, and using \eq{peter} we get $\theta\simeq -0.6$ quite far from
the asymptotic value reported later in Secs.~\ref{best} and \ref{ARA}.
How can we go further and estimate $\theta$ in a safer way? In the
next two sections we shall answer this question.

\section{A good estimate of the lowest droplet exponents}
\label{best}
\begin{figure}[tbp]
\begin{center}
\includegraphics*[height=5.8cm]{thetaeff.eps}
{
\includegraphics*[height=6cm]{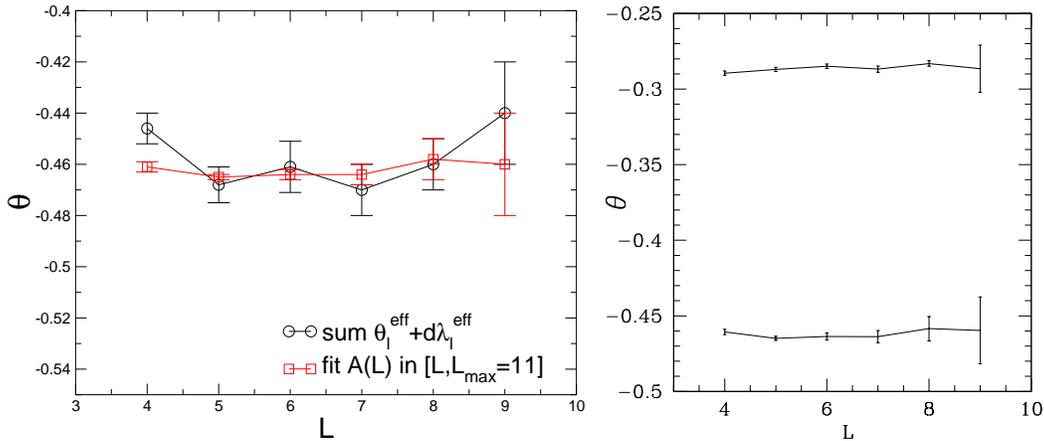}
}

\caption{Exponent $\theta$ for the PPBC case. Left plot: $\theta$
exponent versus $L$ obtained from two methods. Method 1: using
(\ref{a1}). Method 2: using the more stable estimate fitting
(\ref{a2}) over a given range of $L$ values (see text). Right plot:
Domain-wall exponent (top) and $\theta$ exponent (bottom) estimated by
the second method as explained in the text and plotted as a function
of $L$.}
\label{fig4}
\end{center}
\end{figure}

An interesting aspect of the effective $L$-dependent exponents shown
in Figs.~\ref{FIG_lambda} and \ref{FIG_theta} is that, while their FVC
are large, their corrections are of opposite sign. While
$\lambda_l^{\rm eff}(L)$ increases with $L$, $\theta_l^{\rm eff}(L)$
decreases. As they have to be added to get $\theta$ according to the
relation (\ref{general}) their finite-volume corrections cancel out to
a certain degree. If we combine the two estimates for the best data
collapse given in the previous section ($\lambda_l^{\rm eff}=0.70(5)$,
$\theta_l^{\rm eff}\simeq -1.7(1)$) we obtain $\theta\simeq -0.3(2)$
which is very close to the DW value in average. However, this estimate
is too pessimistic. A better route would be to use the two LDEs
estimated from (\ref{lambdaeff}),(\ref{thetaeff}) and adding them according 
to \eq{general}
\be
\theta^{\rm eff}(L)=\theta_l^{\rm eff}(L)+d\lambda_l^{\rm
eff}(L)\; . 
\label{a1}
\ee
In Fig.~\ref{fig4} (left panel) we show the value of $\theta$ obtained
in this way.  Note that the value of the thermal exponent $\theta$ has
{\em negligible} FVC but relatively large statistical fluctuations
with $L$.

A better, albeit related, way to estimate $\theta$ is the following. Instead
of independently finding out $\lambda_l$ and $\theta_l$ we look for an
estimator which depends on the appropriate combination of the
two exponents $\theta=\theta_l+d\lambda_l$. The simplest quantity which
satisfies this requirements is given by the combination, 
\be A(L)=L^d
{\overline{E}(L) \over \overline{v}(L)} \; .  
\label{a2}
\ee 
Since $\overline{E}(L) \simeq L^{\theta_l}$ and $\overline{v}(L)
\simeq L^{d (1-\lambda_l)}$, using (\ref{general}) we obtain $A(L)\sim
L^{\theta}$. To estimate the value of $\theta$ we follow two different
routes: 1) We use (\ref{thetaeff}) by replacing $\theta_l^{\rm
eff}(L)\to\theta^{\rm eff}(L)$ and $\overline{E}(L)\to A(L)$. By
definition, this procedure gives exactly the estimate (\ref{a1}) shown
in the left panel in Fig.~\ref{fig4}. 2) A more stable estimate can be
obtained from a fit of $A(L)$ versus $L$, with data in the range
$[L,\cdots,L_{max}=11]$ (for the PPBC case). This is shown in the left
panel of Fig.~\ref{fig4} together with the previous estimate \eq{a1} and also
in the right panel of Fig.~\ref{fig4} but there compared with the effective
exponent $\theta_{DW}$ obtained from domain-wall calculations.  Our
best value for $\theta$ is
\be
 \theta=-0.46(1)\; .
\label{theta_best}
\ee
This value is very close to the finite-temperature (Monte Carlo or
transfer matrix) estimates $\theta_{TF}=-0.48(1)$ \cite{JAP} but
certainly smaller than the domain-wall value $\theta_{DW}=-0.285$
\cite{VARIOS,RIEGER}. Our estimate for $\theta$ is compatible with the
other possible value $\theta_{TF}$ obtained by other methods as
discussed in Sec.~\ref{intro} but is certainly inconsistent with
the value obtained with other methods with results closer to the
DW estimate.
%
All these estimates strongly support the inequality
$\theta=\theta_{TF}<\theta_{DW}$. However, one cannot exclude a
situation where the present tendency of the data gets modified and
$\theta\to \theta_{DW}$ in the large-$L$ limit~\cite{Moore}. We have
already explained in Sec.~\ref{theta} that $\theta_l$ must converge to
$-2$ in the large volume limit implying the relation
(\ref{general2}). Introducing our estimate (\ref{theta_best}) in
(\ref{general2}) we get,
\be
\lambda_l=0.770(5) \; .
\label{lambda_best}
\ee
A convincing proof of the correctness
of the values (\ref{theta_best},\ref{lambda_best}) requires 
proving that the estimate (\ref{lambdaeff})
converges to the value (\ref{lambda_best}) when $L\to\infty$. In the
next section we present an aspect-ratio analysis to evidence that the
estimates (\ref{theta_best}),(\ref{lambda_best}) are correct in the
large $L$ limit.

\section{Aspect-ratio analysis of the lowest droplet exponents}
\label{ARA}

In this section, we present some additional data obtained via an
aspect-ratio analysis (ARA). This analysis has been proved to be very
useful to extract the value of the domain-wall exponent $\theta_{DW}$
by generating domain walls in rectangular lattices $M\times L$ with
different aspect ratios $M/L$~\cite{CBM02,A02}. It has been found
that, in the limit of large aspect ratio, the value of $\theta_{DW}$
for Gaussian spin glasses is largely independent of the boundary
conditions. We have seen in Sec.~\ref{2d} that our measurements on
squared lattices of size $L\times L$ mix small excitations with large
ones so one does not have a clear-cut separation in the statistical
distribution between the two different regimes $v\sim {\cal O}(1)$ and
$v/V\sim {\cal O}(1)$. Our main motivation here is to show that, by
investigating large aspect-ratios, we can separate these two different
scaling regimes.  We made our measurements on systems of size $L \times M$,
with $M = L R >> L$ where $R$ ranges from $1$ up to $10$. We have
investigated different types of boundary conditions: periodic boundary
conditions in both directions (PPBC) and periodic boundary conditions
in the $L$ direction with free boundary conditions in the $M$
direction (FPBC).

\begin{figure}[tbp]
\begin{center}
\rotatebox{270}{
\includegraphics*[width=8cm,height=10cm]{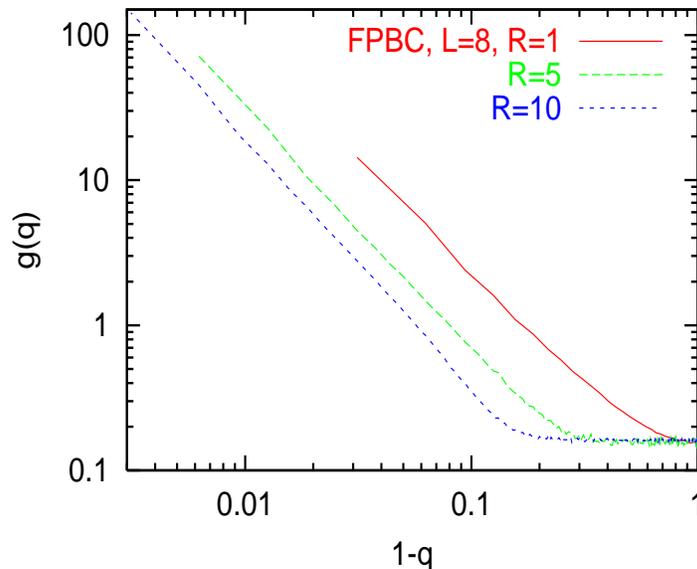}}
\caption{$g(q)$ versus $1-q$ for the FPBC for $R=1, 5$ and $10$ and for
$L=8$.
\label{figure45}}
\end{center}
\end{figure}

\begin{figure}[tbp]
\begin{center}
\rotatebox{270}{
\includegraphics*[width=8cm,height=10cm]{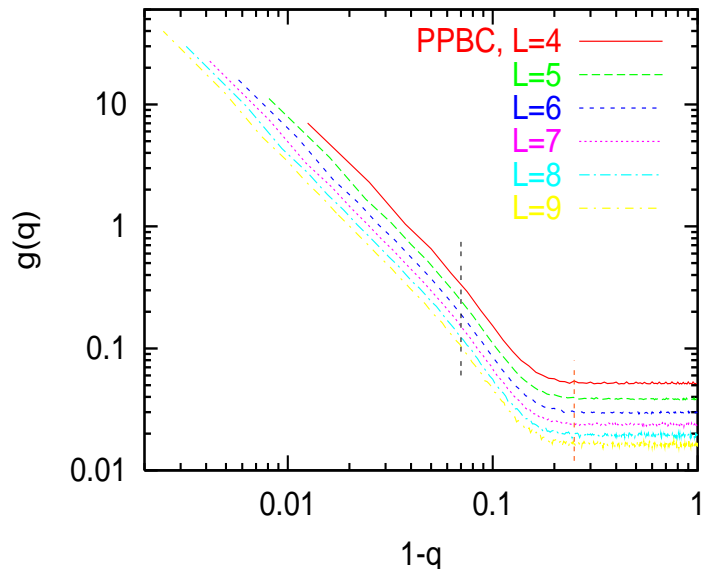}}
\caption{$g(q)$ versus $1-q$ for the PPBC case for $R=10$.
\label{figure46}}
\end{center}
\end{figure}

\begin{figure}[tbp]
\begin{center}
\includegraphics*[width=8cm,height=8cm]{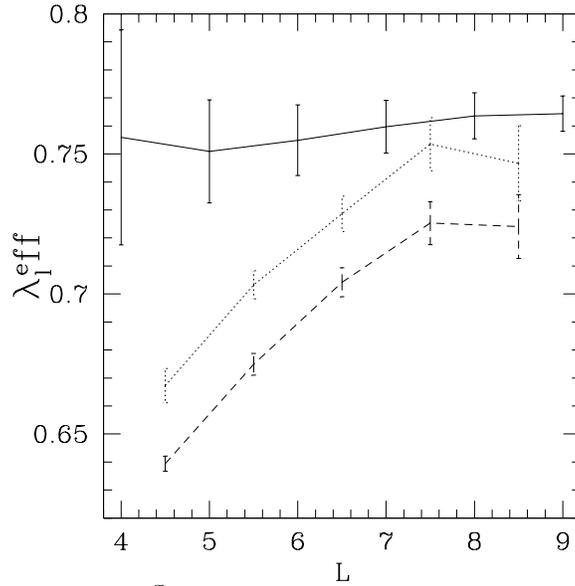}
\caption{Effective lowest droplet exponent $\lambda_l^{\rm eff}$ versus $L$ 
for the PPBC case for $R=10$. We represent the values of $\lambda_l^{\rm eff}$ 
obtained from fitting the distribution of $g(q)$ for small excitations 
(solid line), for large excitations (short dashed line) as well as the value 
obtained by fitting the average size of excitations (dotted line).
\label{figure47}}
\end{center}
\end{figure}

In Fig.~(\ref{figure45}), we display the data for $g(q)$ versus $1-q$
\eq{eq5} for the FPBC case for $L=8$ and $R=1,5$ and $10$. One can
clearly see that the behavior of the distribution $g(q)$ drastically
changes as one increases $R$. Indeed, as we have already seen in
Sec.~\ref{2d} and in \eq{q0},\eq{q1},\eq{eq5}, for $R=1$ it is very
difficult to separate the region of small excitations (a scaling
region with $g(q) \simeq {1 \over (1-q)^{\lambda_l+1}}$) from the one
of large excitations (a constant $q$-independent contribution $g(q)
\simeq { 1 \over V^{\lambda_l}}$).  The main advantage of separating
these two regions is that one can fit directly each of them. This
yields two separate measurements of the LDE $\lambda_l$ in addition to
the estimate \eq{eqv} obtained from the $L$ dependence of the average
size of the excitations.
\vskip -0.15in
In Fig.~(\ref{figure46}), we show $g(q)$ versus $1-q$ for $R=10$ for
various linear sizes $L$ and for the PPBC case. These distributions
have been obtained by running a large number of samples ranging from
10 million of samples for $L=4$ down to 5 million for the largest size
$L=9$.  We have also inserted in the figure two vertical lines which
indicate the limits for the range of values we have chosen for the
fits of the scaling behavior of the finite-size excitation sector
($1-q \leq 0.07$) and for the constant contribution corresponding to
large scale excitations ($1-q \geq 0.25$). We have chosen these values
for the following reasons. First, as one can clearly see in
Figs.~(\ref{figure45},\ref{figure46}), the scaling region for small
excitations survives up to excitation sizes $v \simeq L\times L$. This
size provides a threshold value for the overlap $q_{\rm th}$ below
which the simple scaling $g(q) \simeq {1 \over (1-q)^{\lambda_l+1}}$
does not hold anymore,
\be 1-q_{\rm th}=1-(1-{2 v\over V}) \simeq {2 L^2 \over R L^2}
\simeq {2\over R} \; .  
\ee 
An second, there is a crossover region around $q \simeq q_{\rm
th}$. A careful look at Fig.~(\ref{figure46}) shows that the scaling
region for small excitations ends around $1-q \simeq 0.07$. At this
value, one observes a change of the slope of the curves just before
entering the regime of large excitations where $g(q)$ becomes $q$
independent. For $1-q \ge 0.25$, the curves are rather constant and
the result of a fit does not depend much on the choice
$1-q=0.25$. This second threshold value is indicated as the rightmost
vertical bar in Fig.~(\ref{figure46}).
\vskip -0.15in
In Fig.~(\ref{figure47}), we show the estimated values of effective
lowest droplet exponent $\lambda_l^{\rm eff}$ obtained in three different
ways. The first estimate has been obtained by averaging the volume of
all excitations for different lattice sizes as explained in
Sec.~\ref{2d} and then taking a logarithmic derivative, see
\eq{lambdaeff}.  The second estimate has been obtained by considering 
the large excitation sector ($1-q \ge 0.25$) and its $L,R$ dependence:
\be 
g(q) \simeq (RL^2)^{-\lambda_l} \; .  
\ee 
Averaging the excitation volume within this sector ($1-q \ge 0.25$)
and using again the corresponding logarithmic derivatives as in
\eq{lambdaeff} yields the second estimate. The third estimate for
$\lambda_l$ is obtained from a direct fit of $g(q)$ for small values
of $1-q$ : 
\be 
g(q) \simeq (1-q)^{-1-\lambda_l} \; .
\ee
This third method is in fact the most direct one since it can be done
for each size $L$ (while the other two estimates require a fit using data from two
different lattice sizes $L$ and $L'$). The first conclusion that we learn from
Fig.~(\ref{figure47}) is that the ARA produces a great improvement on
the estimated values of the exponent $\lambda_l$. The most stable
measurement is the third estimate obtained by fitting the small-size
spectrum of the excitations. In that case, $\lambda_l^{\rm eff}$ is nearly
constant with a value that converges to

\be
\lambda_l=0.77(1) 
\label{lambda_ARA}
\ee
in excellent agreement with the result \eq{lambda_best} of the
previous section.
\vskip -0.15in
Moreover, one also observes in Fig.~(\ref{figure47}) that the two
other estimated values for $\lambda_l^{\rm eff}$, obtained with the first and
second methods, are strongly correlated. This shows that finite-volume
corrections, which are expected to affect the value of the exponent
obtained from the analysis of large-size excitations, does affect also
the value of the exponent obtained by averaging over the whole
spectrum. In addition, we also observe that the ARA for large $R$
strongly decreases the magnitude of finite-volume corrections. While
on a square geometry, the effective exponent $\lambda_l^{\rm eff}$ obtained from
the average size of excitations took values in the range $0.52-0.62$
(see Fig.(\ref{FIG_lambda})), with the ARA, we obtain for the same
exponent values in the range $0.64-0.72$, which are much closer to the
expected asymptotic value $0.77(1)$.

\begin{figure}[tbp]
\begin{center}
\includegraphics*[width=10cm,height=8cm]{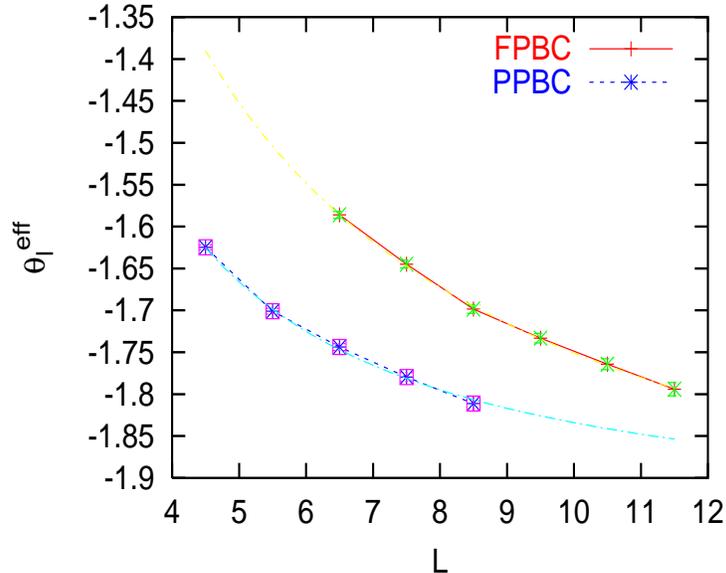}
\caption{Effective exponent $\theta_l^{\rm eff}$ obtained via a 
logarithmic derivative for the PPBC and the FPBC. We also plot 
best fit curves which converge to $\theta_l^{\rm eff}(L\rightarrow \infty)= 
-1.96(6)$ for the PPBC and to $\theta_l^{\rm eff}(L\rightarrow \infty)= 
-2.12(11)$ for the FPBC.
\label{figure48}}
\end{center}
\end{figure}
\vskip -0.15in

The same conclusion holds for the lowest droplet exponent $\theta_l$.
In Fig.~(\ref{figure48}), we show the effective exponent
$\theta_l^{\rm eff}$ obtained by evaluating the logarithmic derivative
as in \eq{thetaeff}. Note that finite-volume corrections are much
smaller than with the squared lattices and as a result, the value
of the effective exponent converges much faster to the expected value
$-2$. Using a fit of the form $\theta_l^{\rm eff}(L)=\theta_l^{\rm
eff}(\infty)+{cst\over L^\alpha}$, one gets $\theta_l^{\rm
eff}(\infty)=-1.96(6)$ for the PPBC, the best fit being also
represented in Fig.~(\ref{figure48}). In this figure, we also show the
same exponent obtained for the FPBC, where the best fit yields the
asymptotic value $\theta_l^{\rm eff}(\infty)=-2.12(11)$. In both cases
the fitting value we obtain for the exponent is $\alpha\simeq 1$. Note that the
asymptotic values for $\theta_l^{\rm eff}$ are well compatible with
our prediction of Sec.~\ref{theta}, $\theta_l=-2$ (see also the
heuristic argument in Appendix \ref{upperbound}).

\section{Finite-volume corrections (FVC) and their relation to 
the statistics of extreme values}
\label{fse}

What is the origin of these strong finite-volume corrections?
Intuitively it is not difficult to find an explanation for the strong
systematic finite-volume corrections in the lowest droplet exponent
$\theta_l$. As the word {\em lowest} indicates, these exponents
describe the statistical distribution of droplet excitations which are
at the tail of the energy gap distribution that includes all possible
high energy levels. As the volume of the system increases there is
more available space to find excitations with lower energy gap. This
implies that there is more probability to find a lowest droplet with
an energy smaller than a given threshold value $E^*$. As this
probability systematically increases with $L$ for all samples,
$\theta_l^{\rm eff}(L)$ must be a decreasing function of $L$. This is
in agreement with what we have found. The behavior of $\lambda_l^{\rm
eff}(L)$ is more difficult to establish.  By the same token, although
there is more available space for droplets we expect that different
droplet sizes increase or decrease their relative probability in a
non-trivial way making difficult to guess how the exponent
$\lambda_l^{\rm eff}(L)$ systematically changes with $L$.
\begin{figure}[tbp]
\begin{center}
\includegraphics*[height=0.4\columnwidth]{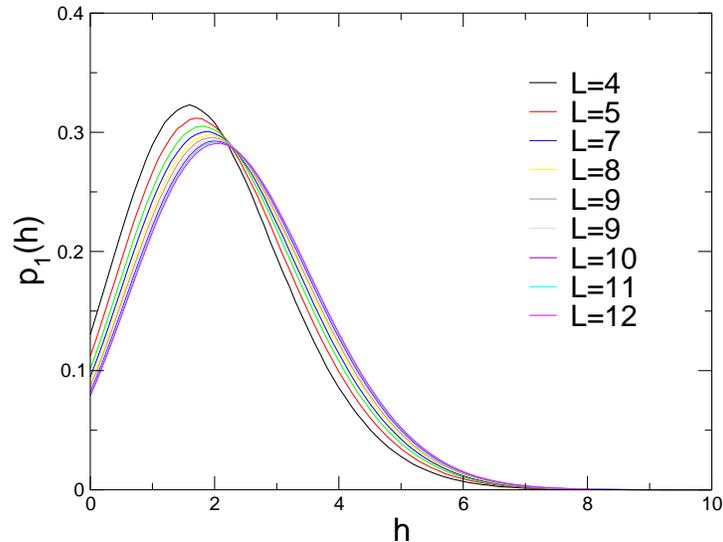}
\caption{Local-field distribution for different lattice sizes with 
FFBC boundary conditions.
\label{local_field}}
\end{center}
\end{figure}

To understand the origin of finite-volume corrections in the value of
$\theta_l$ we have focused our attention on the behavior of the upper
bound exponent $\theta_l^1$ describing the statistics of the lowest
one-spin excitations as described in the
Appendix~\ref{upperbound}. The gap distribution corresponding to these
excitations can be obtained from the local-field distribution
evaluated at the ground state. We have numerically computed this
distribution for different sizes, the results are shown in
Fig.~\ref{local_field}. As discussed in the appendix~\ref{upperbound},
the local-field distribution has a finite weight at zero field and is
a self-averaging quantity. As the local field distribution is
self-averaging, the probability distribution for the lowest one-spin
excitations corresponds to the extreme value statistics of the
local-field distribution $p_1(h)$ where $h$ stands for the local field
which we assume to be positive as the gap is given by its absolute
value (the subindex 1 is used to stress that this distribution
describes energy gaps for one-spin excitations only). If $P_1(h)$
stands for the probability distribution of the smallest local-fields,
then $P_1(h)$ can be easily related to $p_1(h)$ by standard probability
arguments (see for instance, \cite{MB}). Although the argument is very
general, here we apply it to one-spin excitations. For a given sample,
the lowest value $h$ is selected as the minimum value among all the
possible $V$ local fields $h_i$ at each lattice site. The probability
$P_1(h)$ is given by the following expression,
\be
P_1(h)=Vp_1(h)(1-\int_{h}^{\infty}p_1(h')dh')^{V-1}=-\frac{\partial} {\partial h}
\Biggl( \int_{h}^{\infty} p(h')dh'\Biggr)^V
\label{fse1}
\ee
which accounts for all possible ways the value $h$ coincides with the
minimum value obtained among all different $V$ local fields distributed
according to the $p_1(h)$. The last identity shows that $P_1(h)$ is
normalized. This probability can be explicitly worked out in the large
$V$ limit,
\be
P_1(h)=-\frac{\partial}{\partial
h} \exp\left[ -Vg_1(h)\right]=Vg_1(h)\exp\left[ -Vg_1(h)\right] \; .
\label{fse2}
\ee
Up to second order in $h$ the function $g_1(h)$ is given by,
\be
g_1(h)=p_1(0)h+\frac{p_1'(0)+(p_1(0))^2}{2}h^2 \; .
\label{fse3}
\ee
From (\ref{fse2}) we immediately learn that the gap distribution is an
exponential with a sub-leading Gaussian correction whose magnitude
decreases as $1/V$. Actually, plotting $P_1(h)/V$ as function
of the scaling variable $x=hV$ one gets,
\be
\frac{P_1(h)}{V}=g_1'(x/V)\exp\left[ -xp_1(0)
-\frac{p_1'(0)+(p_1(0))^2}{2V}x^2\right] \; .
\label{fse4}
\ee
In the large $V$ limit $g_1'(x/V)\to p_1(0)$ and the coefficient in front of
the Gaussian correction goes asymptotically to zero, therefore the
distribution $P_1(h)$ converges to an exponential as expected,
\be
P_1(h)=Vp_1(0)\exp\bigl( -Vp_1(0)h\bigr)
\label{fse5}
\ee
in agreement with the scaling relation (\ref{eq7}).  We can now
understand the deviations from the pure exponential behavior discussed
in Sec.~\ref{2d} in the context of the inset a) shown in
Fig.~\ref{FIG2}. They are simply consequence of the finite-volume
corrections of the extreme values of the gap distribution for all
energy levels (and not only one-spin excitations as we are discussing
here). So one could imagine to compute for a given sample the first
$V$ energy gaps corresponding to the first $V$ excitations. We also
assume that the resulting distribution $p_{\rm all}(E)$ is
self-averaging in the large volume limit (as it is the local-field
distribution $p_1(h)$). The lowest energy distribution constructed by
taking the minimum value of the gap $E$ for each sample yields the
extremes distribution $P_{\rm all}(E)$ defined in (\ref{eq7}) (from
now on we will drop the subindex '{\rm all}' in $P_{\rm all}(E)$ as it
coincides with the $P(E)$ defined in (\ref{eq7}). Also the subindex
'all' for the $p_{\rm all}(E)$ will be dropped). In fact, the
parameters $p(0),p'(0)$ characterizing this
distribution can be obtained from the $P(E)$'s shown in
Fig.~\ref{FIG2}.  To evaluate them, the best way is to analyze the
cumulative distribution ${\cal P}(E)=\int_{E}^\infty d\,E'P(E')$ which
from \eq{fse1} we can assume to be $ {\cal P}(E)=\exp[-V\,g_{\rm
all}(E)]$. Thus we can fit ${\cal P}(E)$ with an exponential with
Gaussian corrections $A\exp[-Bx-Cx^2/2]$ whose fitting parameters are
related to $p(0)$ and $p'(0)$. The best fits yield
the following values $p(0)\approx 0.2$ and $p'(0)=0.3$.

\begin{figure}[tbp]
\begin{center}
\includegraphics*[height=0.4\columnwidth]{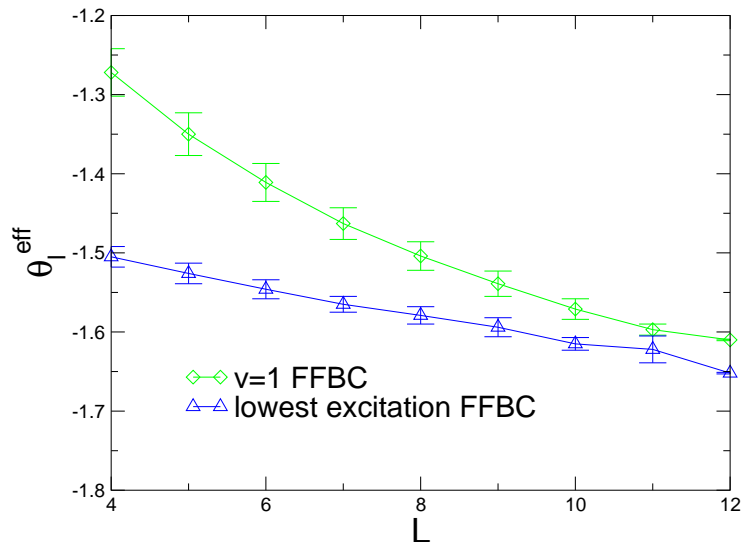}
\caption{Effective droplet exponent $\theta^{\rm eff}_l$ versus $L$ for the 
FFBC  case, computed using logarithmic derivatives (see text). We show the 
exponent obtained  for one-spin excitations ($v=1$) in comparison to the 
one obtained from the whole distribution of gaps.
\label{FIG_theta1}}
\end{center}
\end{figure}
\vskip -0.15in

Coming back to our original goal we discuss now the finite-volume
corrections for the estimate $\theta_l^{\rm eff}$, as shown in
Fig.~\ref{FIG_theta}. From the distribution (\ref{fse2}) describing
 the whole spectrum of excitations we can express the effective exponent
(\ref{thetaeff}) for $L>>1$ as,
\be
\theta_l^{\rm eff}(L)=\frac{\partial
\log\bigl(\overline{E}(L)\bigr)}{\log(L)} \; .
\label{fse6}
\ee
The computation of $\overline{E}(L)$ is quite straightforward as it is
given by the simple relation,
\be
\overline{E}(L)=\int_0^{\infty}EP(E)dE=\int_0^{\infty}\exp\Bigr(
-Vg(E)\Bigl)
\label{fse7}
\ee
where we have used (\ref{fse2}) plus an integration by parts. The
integral, up to second order in $1/V$ yields,
\be
\overline{E}=\frac{1}{Vp(0)}-\frac{p'(0)+(p(0))^2}{V^2(p(0))^3}
+O\Bigl(\frac{1}{V^3}\Bigr) \; .
\label{fse8}
\ee
Inserting this result in (\ref{fse6}) we finally get,
\be
\theta_l^{\rm eff}(L)=-d+\frac{d}{V}\bigl(1+\frac{p'(0)}{(p(0))^2}\bigr)
+O\Bigl(\frac{1}{V^3}\Bigr) \; .
\label{fse9}
\ee
This shows that $\theta_l^{\rm eff}(L)$ approaches $-d$ from below (as
$p'(0)$ is positive). On the other hand the magnitude of the
finite-volume corrections can be pretty large if
$\frac{p'(0)}{(p(0))^2}>>1$. For instance, if one takes the results obtained
from the analysis of one-spin excitations one gets $p_1(0)\simeq
0.069,p_1'(0)\simeq .125$ yielding $\frac{p_1'(0)}{(p_1(0))^2}\simeq 27.$ which is
indeed large. Inserting these values in (\ref{fse9}) we obtain an estimation 
for $\theta^{\rm eff}_l(L=12)=-1.65$ in good agreement with numerical results 
(see Fig.~\ref{FIG_theta1}).

If we insert the previous estimated values for the whole spectrum of
excitations extracted from the $P(E)$'s in Fig.~\ref{FIG2}, we obtain
$\frac{p'(0)}{(p(0))^2}\simeq 7.5$. From (\ref{fse9}) it follows that
$\theta_l^{\rm eff}(L)\simeq -2(1-7.5/V)$, which for $L=11$ yields
$\theta_l^{\rm eff}=-1.87$. All in all, the magnitude of the effective
exponent $\theta_l$ is well compatible with the reported value
$\theta_l^{\rm eff}$ used in the inset a) in Fig.~\ref{FIG2} for the
PP case. Note that the FVC corrections to $\theta^{\rm eff}_l$
obtained from the local-field distributions in the FF case are much
larger than FVC corrections in the PP case in agreement with ARA
results (see Fig.~\ref{figure48}). From this analysis it becomes clear
that to significantly reduce the magnitude of the finite-volume
corrections in the value of $\theta_l$ (let us say $\theta_l\simeq
-1.95$), we would need larger volumes beyond $20\times 20$.

\section{Compactness of the lowest energy droplets.}
\label{compact}

One intriguing question about the droplet excitations concerns their
topological properties. Kawashima and Aoki~\cite{KA} have argued that
droplet excitations are not compact. Instead, their volume has a
fractal structure as the number of lattice points included in the
droplet scales with its spanning length (which is a measure of the
length scale of the droplet) with an exponent smaller (around 1.80(2))
than the dimension of the system (2).

\begin{figure}[tbp]
\begin{center}
\includegraphics*[height=0.4\columnwidth]{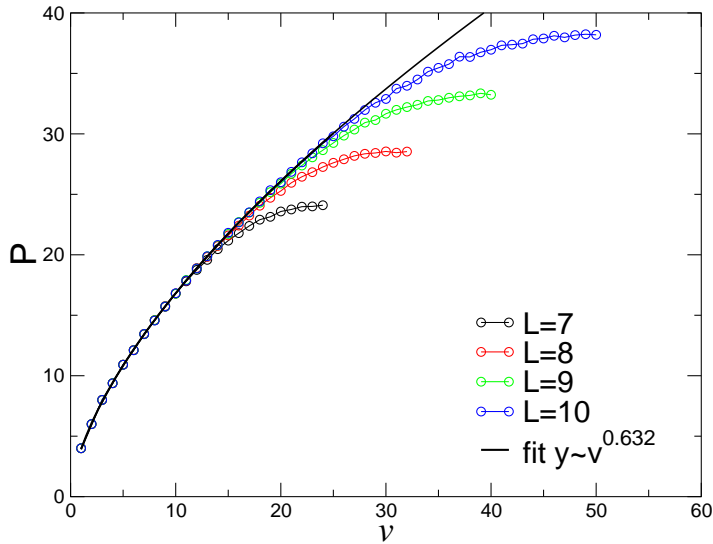}
\caption{Perimeter ($P$) of the droplet versus its volume ($v$). 
The solid line corresponds to the fit (\ref{Pv}) with $d_s/d_v=0.632(2)$.} 
\label{FIG_DF}
\end{center}
\end{figure}

To answer this question we have computed the surface, i.e. the
perimeter $P$, of all lowest droplets. The relation between the
average perimeter as function of the size $v$ of the excitation
depends on both the fractal dimension of the surface or perimeter
$d_s$ and the volume $d_v$ of the lowest droplets. $d_s$ and $d_v$ can
be defined in terms of the spanning length $l$ of the droplet which
can be defined in different ways. For example, one could use the
gyration radius, the average distance between the sites contained in
the cluster, or the maximum distance among the sites of the
cluster. As the typical length scale of our lowest droplets is small,
$l\simeq 10$, we have not attempted to estimate it as this can
strongly depend on the precise definition of the spanning length.
Here, we restrict ourselves to investigate the perimeter-volume dependence.
In terms of the spanning length $l$ the surface fractal and
volume fractal dimensions $d_s,d_v$ of the droplets are defined as,
\bea
l\sim P^{\frac{1}{d_s}}\label{ds}\\
l\sim v^{\frac{1}{d_v}}
\label{dv}
\eea
which combined give,
\be
P \sim v^{\frac{d_s}{d_v}}\; .
\label{Pv}
\ee
In Fig.~\ref{FIG_DF} we show $P(v)$ as a function of $v$ for different
lattice sizes. As can be seen, FVC are important for large
volumes. However, there is an enveloping curve that is independent of
$L$ for small volumes and spans a progressively increase range of
volumes as $L$ increases. This enveloping curve is excellently fitted
(continuous curve) by the scaling relation (\ref{Pv}) and yields an estimate,
\be
\frac{d_s}{d_v}\simeq 0.632(2) 
\label{dvds}
\ee
consistent with the results reported by Kawashima and Aoki
$\frac{d_s}{d_v}=0.61(1)$ obtained with a completely different method.

\vskip -0.15in

\section{Conclusions}
\label{conclusions}

We have shown that a proper description of low-temperature properties
in two-dimensional Gaussian spin glasses can be done in terms of the
lowest droplet exponents (LDEs) $\lambda_l$ and $\theta_l$ describing
the spectrum of lowest excitations. $\lambda_l$ describes the spectrum
of sizes of the lowest energy droplets, while $\theta_l$ describes the
typical energy cost of these lowest droplets whatever their
size. Assuming that $\theta_l=-d$ one concludes that the LDE $\lambda_l$ 
fully characterizes the spin-glass phase. Although
independent numerical estimates of $\theta_l$ and $\lambda_l$ show
strong finite-volume corrections, the thermal exponent
$\theta=\theta_l+d\lambda_l$ can be well estimated giving the results
\eq{theta_best},\eq{lambda_best}
\be
 \theta=-0.46(1)~~~~~\lambda_l=0.770(5).
\label{best_estimates}
\ee
unambiguously showing that $\theta< \theta_{DW}=-0.287(4)$
\cite{A02}. Our estimates \eq{best_estimates} have been confirmed via
an aspect-ratio analysis which provides estimates much less influenced
by finite-volume corrections.  Moreover, the result $\theta_l=-2$
(that is believed to be correct for spin glasses with coupling
distributions with finite weight at zero coupling, see the
Appendix~\ref{upperbound}) has been numerically confirmed by the
aspect-ratio analysis. To sum up, McMillan's excitations are not the typical
low-lying excitations and our approach offers a new and independent way
to estimate the thermal exponent $\theta$ without the need to generate
typical low-lying excitations by looking at the new ground state of the
system after perturbing it.

We think that discrepancies on the value of the thermal exponent
$\theta$ reported by comparing {\em non-perturbative} methods (such as
finite-temperature transfer-matrix calculations and the present lowest
droplet analysis) with {\em perturbative methods} such as
domain-wall calculations (or perturbations induced by introducing a
coupling term in the energy function that induces a large-scale
excitation) are serious enough to be taken as a clear indication that
our knowledge of the low-temperature properties of the 2d GISG is
still inadequate.  In this direction we want also to recall the issue of
multifractality and the possibility that different exponents could
describe the zero-temperature critical point. Is this really possible?
Well, to our knowledge no exact result precludes this possibility and,
although purely speculative at the present stage, one should seriously
think about it. Altogether, the present analysis suggests that the
excitations in 2d GISG are very different from the compact droplets
proposed in the context of the droplet model. If this were true, the
implications of the 2d studies in larger dimensions could be
important.  There are many routes that can be followed to understand
better what is going on and the origin of this discrepancy. Certainly,
with the outstanding accuracy of present algorithms to compute ground
states in 2d, it would be very interesting to revisit again the
analysis of the statistics of the large scale excitations generated by
imposing a uniform magnetic field. ``Old'' results by Rieger et
al.~\cite{RIEGER} give an estimate for $\theta$ that is compatible
with our estimate rather than to the domain-wall estimate. This would
be an independent check of our values, but using a perturbation method
with an appropriate {\em neutral observable} such as the global
magnetization as has been explained in Sec.~\ref{intro} before \eq{eqP}.
 
The proposed method may appear venturesome as, to our present
knowledge, there is no numerical study in the field along this line of
research.  However, as explained in Sec.~\ref{intro}, recent studies
on the disordered Anderson model~\cite{GEFEN} have revealed that the
analysis of the lowest excitation provides a good description of the
localized phase. More studies are certainly required to understand
better the reliability of the present method to investigate the
critical properties of spin glasses. One disadvantage of our approach
is that a huge number of samples is needed to reasonably sample
large-scale excitations. However, as we saw in Sec.~\ref{ARA}, the
behavior of the $g(q)$ for small $1-q$ can be extracted with a modest
number of samples. The advantage, as has been already stressed in
Sec.~\ref{intro}, is that we do not introduce any external
perturbation to generate the excitations.

Finally, we want to comment on the extension of this approach to other
models. Of course, the immediate extension one could think of is the
2d $\pm J$ model. However, the analysis of this model appears quite
troublesome. This model does not have a continuous gap distribution
but a discrete one that introduces further complications. As the
ground state is not unique one has to redefine the full analysis to
properly define the spectrum of lowest excitations. The discreteness
of variables could have some unexpected effects in the present
approach as seems to happen also with domain-wall
calculations~\cite{HY01,A02}. It is more natural to extend the
research to other models such as 2d ISG with other continuous coupling
distributions without gap (e.g. characterized by $P(J)\sim
|J|^{\alpha}$ for $|J|\to 0$), Migdal-Kadanoff spin glasses (where
both the ground state and the first excitation could be feasibly found
with an appropriate algorithm), Gaussian spin glasses beyond $d=2$
(where unfortunately, algorithms are much less effective than in 2d as
the finding of the ground state becomes a NP complete problem) and
finally mean-field spin-glass models where the zero temperature
exponents are known and maybe the spectrum of lowest excitations could
be analytically tackled. Preliminary results in this case~\cite{MF2}
confirm that the present analysis describes pretty well the data for
rather small sizes. We are pretty confident that, in the near future,
new results and evidences will finally resolve this interesting
problem.

\appendix
\section{Heuristic proof of the identity $\theta_l=-d$}
\label{upperbound}

In this appendix we show that $\theta_l=-d$. In what follows we do not
attempt to present a rigorous proof but we content ourselves to
present an heuristic argument. The argument has two parts: first we
show that $-d$ is an upper bound, next we show that the upper bound is
the exact value. For the upper bound the argument is well known and
goes as follows. Consider the ground state and all possible one-spin
excitations. Because one-spin excitations are not necessarily the
absolute lowest ones, the statistics of the lowest one-spin
excitations must yield an upper bound $\theta_l^{1}$ for the value of
$\theta_l$, $\theta_l\le \theta_l^{1}$. The statistics of the lowest
one-spin excitations is determined by the behavior of the ground-state
local field distribution $p(h)$ in the limit $h\to 0$. If $p(h)$ is
self-averaging and $p(0)$ is finite (in the large-volume limit) then
the statistics of the lowest excitations must be governed by the
exponent $\theta_l^{1}=-d$. Although we do not know a precise
mathematical proof of the statement that $p(0)$ is finite, it looks
quite intuitive~\cite{HUSE}. In any short-range system with a
frustrated ground state and a coupling distribution with finite
density at zero coupling, we may expect a finite probability to find a
cage containing a spin coupled to its neighbors by a set of weak bonds
which produce a vanishing net local-field acting on that spin. This
argument should generally hold for $d\ge 2$. Moreover, as its name
indicates, the local-field distribution is a local observable. An
argument {\em \`a la Brout} proves that it should be self-averaging as
all possible local field values are realized across the whole
lattice~\footnote{Our numerical results in the 2d GISG confirm this
conclusion, see Sec.~\protect\ref{fse}}. The next part of the argument
consists in proving that an identical upper bound is valid by
considering excitations with size strictly larger than 1 but
finite. The upper bound derived for the one-spin excitations must
necessarily hold for finite-size excitations beyond one-spin
excitations (for instance two-spins, three-spins and so on) as the gap
corresponding to the finite-size excitations can always be written as
a linear combination of a finite number of local fields with
coefficients which depend on the ground state configuration. It is
easy to verify that the aforementioned properties of the local-field
distribution $p(h)$ imply that the new gap distribution has a finite
weight at zero gap and is self-averaging. This argument however cannot
be extended to large-scale (with $v\sim V$) excitations in a
straightforward way because the distribution for the corresponding gap
distribution corresponds to an infinite sum of terms in the
$V\to\infty$ limit. However once we argue that $\theta_l$ is an upper
bound valid for all finite-size excitations it can be concluded that
this upper bound must coincide with the exponent $\theta_l$ describing
the probability of the absolute lowest excitations. From~(\ref{eq4})
the fraction of large scale excitations $v\to V$ is given by
$Vg_V=\frac{1}{V^{\lambda}}$. In general $\lambda>0$ so this fraction
vanishes\footnote{This fractions is finite only in $d=1$ where $\lambda=0$. 
But this case is trivial as the surface of large scale
droplets in $d=1$ only contains a finite number of broken bonds.} in
the infinite-volume limit and finite-size excitations determine the
result $\theta_l=-d$ as they dominate the spectrum of lowest
excitations. Moreover, if large-scale excitations yield a different
value for $\theta_l$ this would imply that boundary conditions could
affect the value of the thermal exponent. That would be quite unusual
as this would mean that the exponents of the $T=0$ fixed point would
depend on the boundary conditions.

\section{Transfer Matrix algorithms}
\label{transfermatrix}
In this appendix, we will briefly explain how we determine the ground
state and the first excited state. We will work on a square lattice
of size $L \times L$. The energy associate to a configuration of spins 
$S(i,j)$ with a fixed configuration 
of disorder $J^x(i,j)$ and $J^y(i,j)$ is
\bea
E=\sum_{i=1,L-1} \sum_{j=1,L} J^x(i,j) S(i,j)S(i+1,j) 
+ \sum_{i=1,L} \sum_{j=1,L-1} J^y(i,j) S(i,j)S(i,j+1) \\
+ B_1 \sum_{j=1,L} J^x(L,j) S(L,j)S(1,j) 
+ B_2 \sum_{i=1,L} J^y(i,L) S(i,L)S(i,1)\; , \nonumber
\eea
where $B_1$ and $B_2$ correspond to the choice of boundary conditions.
Here we will consider three cases~: Periodic-Periodic
boundary conditions (PPBC) with $B_1=B_2=1$; Free-Periodic boundary
conditions (FPBC) with $B_1=1 ; B_2=0$ (or equivalently $B_1=0 ;
B_2=1$) and Free-Free boundary conditions (FFBC) $B_1=B_2=0$. We will
only consider the case with a Gaussian distribution of the bond
disorder $J^x, J^y$. To determine the ground state and the first
excited states, we proceed as follow: we start by associating a weight
for each configurations of spins in the first row of the lattice
$S(1,1),S(1,2),\cdots,S(1,L)$:
\be
\label{tfm2}
W(S(1,1),S(1,2),\cdots,S(1,L)) = B_2 J^y(1,L)S(1,L) S(1,1) 
+ \sum_{i=1,L-1} J^y(1,i) S(1,i)S(1,i+1) \; .
\ee
Next, we start iterating the transfer matrix using a sparse-matrix 
factorization \cite{nightingale}. The first iteration gives 
\be
\label{tfm3}
W^1(S(1,2),\cdots,S(1,L),S(2,1)) = max_{S(1,1)}\left[J^x(1,1) S(1,1) S(2,1) 
+ W(S(1,1),\cdots,S(1,L))\right] \; .
\ee
Since we are also interested in the first excited state, we define 
the second largest weight:
\be
\label{tfm4}
W^2(S(1,2),\cdots,S(1,L),S(2,1)) = min_{S(1,1)}\left[J^x(1,1) S(1,1) S(2,1) 
+ W(S(1,1),\cdots,S(1,L))\right]\; .
\ee
In the following, we will use the simplified notation 
\be
W(i,j)\equiv W(S(i,j),\cdots,S(i,L),S(i+1,1),\cdots,S(i+1,j-1)) \; .
\ee
Thus eqs.(\ref{tfm3},\ref{tfm4}) become 
\be
W^1(1,2) = max_{S(1,1)}\left[J^x(1,1) S(1,1) S(2,1) + W(1,1)\right]
\ee
and 
\be
W^2(1,2) = min_{S(1,1)}\left[J^x(1,1) S(1,1) S(2,1) + W(1,1)\right]  \; .
\ee
At the next iteration, the numbers of possible weight will again be
multiplied by two but we will keep only the two largest ones defined
as
\bea
\label{tfm39}
W^1(1,3)= max_{S(1,2)}\left[J^x(1,2) S(1,2) S(2,2)  +  J^y(2,1) S(2,1)S(2,2) 
+ W^1(1,2)\right]
\eea
\bea
\label{tfm40}
 W^2(1,3)&=& max(min_{S(1,2)}[J^x(1,2) S(1,2) S(2,2)  
+  J^y(2,1) S(2,1)S(2,2) + W^1(1,2)], \\
&&\; \; \; \; \; \; \; \; max_{S(1,2)}[J^x(1,2) S(1,2) S(2,2)  
+  J^y(2,1) S(2,1)S(2,2) + W^2(1,2)] )\; .\nonumber 
\eea
The general iteration relations are
\bea
\label{tfm41}
W^1(i,j)&=& max_{S(i,j-1)}[J^x(i,j-1) S(i,j-1) S(i+1,j-1)  \\
&&\; \; \; \; \; \; \; \;  \; \;  \; \;  \;  \; \; \; \; \; \; 
 +  J^y(i+1,j-2) S(i+1,j-2)S(i+1,j-1) + W^1(i,j-1)]\nonumber
\eea
\bea
 W^2(i,j)= max&(&min_{S(i,j-1)}[J^x(i,j-1) S(i,j-1) S(i+1,j-1)   \\
&&  \; \; \; \; \; \; \; \;  \; \;  \; \; \;  \; \; \; \; \; 
+  J^y(i+1,j-2) S(i+1,j-2)S(i+1,j-1) + W^1(i,j-1)], \nonumber\\
&&  max_{S(i,j-1)}[J^x(i,j-1) S(i,j-1) S(i+1,j-1)  \nonumber\\
&& \; \; \; \; \; \; \; \;  \; \;  \; \; \;  \; \; \; \; \; 
+  J^y(i+1,j-2) S(i+1,j-2)S(i+1,j-1) + W^2(i,j-1)]) \; . \nonumber 
\eea
In addition, each time that we end the construction of a new row, 
we must add the boundary term
\be
W^1(i,1) \rightarrow W^1(i,1)  + B_2 J^y(i,L)S(i,L)S(i,1)\; ; \;   
W^2(i,1) \rightarrow W^2(i,1)  + B_2 J^y(i,L)S(i,L)S(i,1)   \; .
\ee

We still have to take in account the boundary condition corresponding
to $B_1$. The two type of boundary conditions (Free or Periodic) 
have to be considered separately :
\begin{itemize}
\item Free boundary condition, $B_1=0$ : In that case, we iterate up to the
construction of the weights associated to the configurations of the
spins $S(L,1),\cdots,S(L,L)$. The energy of the ground state ($E_0$)
is then simply the maximum among all the weights $W^1(L,1)$:
\be
E_0 = max_{\{S(L,1),\cdots,S(L,L)\}}[W^1(L,1)].
\ee
We call $\{S^0(L,1),\cdots,S^0(L,L)\}$ the configuration of spins on
the last row for the ground state. The energy of the first excited
state is the second largest weight among $W^1(L,1)$ and $W^2(L,1)$:
\be
E_1 = max[max_{\{S(L,1),\cdots,S(L,L)\}\neq \{S^0(L,1),\cdots,S^0(L,L)\}} 
[W^1(L,1)],max_{\{S(L,1),\cdots,S(L,L)\}}[W^2(L,1)]] \; .
\ee

\item Periodic boundary case, $B_1=1$ : We first choose one configuration of
spins on the first row $S^i(1,1),S^i(1,2),\cdots,S^i(1,L)$. The weight
of this configuration is defined as in (\ref{tfm2}).  The weight of
all the other configurations of spins on this first row are fixed to
an arbitrary large negative number. Next we iterate the transfer
matrix as described above, up to the construction of the weights
$W^1(L,1)$ and $W^2(L,1)$. Finally, we iterate one additional row,
with bonds $J^x(L,i)$ and $J^y(L+1,i)=0$. Next, we store the two
weights $W^1$ and $W^2$ associated to the initial spins configuration
$S^i(1,1),S^i(1,2),\cdots,S^i(1,L)$. We denote these two weights by
\bea
{\cal W}^1&& (S^i(1,1),S^i(1,2),\cdots,S^i(1,L))  \equiv  \\
&&W^1(S(L+1,1)=S^i(1,1),S(L+1,2)=S^i(1,2),\cdots,S(L+1,L)=S^i(1,L)) 
\nonumber \\
{\cal W}^2 &&(S^i(1,1),S^i(1,2),\cdots,S^i(1,L))  \equiv  \\
&& W^2(S(L+1,1)=S^i(1,1),S(L+1,2)=S^i(1,2),\cdots,S(L+1,L)=S^i(1,L))\nonumber 
 \; . 
\eea

The energy of the ground state is the maximum on all the ${\cal W}^1$:
\be
E_0 = max_{\{S^i(1,1),\cdots,S^i(1,L)\}}[{\cal W}^1(S^i(1,1),S^i(1,2),
\cdots,S^i(1,L))] \; ,
\ee
and we denote by $S^0(1,1),\cdots,S^0(1,L)$ the configuration of spins
on the first row for the ground state. The energy of the first excited
state is the second largest weight among ${\cal W}^1(L,1)$ and ${\cal
W}^2(L,1)$:
\bea
E_1 = max[max_{\{S^i(1,1),\cdots,S^i(1,L)\}\neq \{S^0(1,1),\cdots,S^0(1,L)\}}
[{\cal W}^1(S^i(1,1),S^i(1,2),\cdots,S^i(1,L))],\\
max_{\{S^i(1,1),\cdots,S^i(1,L)\}}[{\cal W}^2(S^i(1,1),S^i(1,2),
\cdots,S^i(1,L))]] \; .
\eea
The construction of the ground state and of the first excited state 
is much more costly in computing time for the periodic case, 
since we have to repeat $2^L$ times the iterations, one time for each 
configuration $S^i(1,1),S^i(1,2),\cdots,S^i(1,L)$. 
\end{itemize}

So far, we have only described how to compute the value of the
energies associated to the ground state and the first excited
state. Since we also want to determine the spins configurations for
these two states, we have to store, at each iteration of the transfer
matrix, the value of the spin on which one sums, as well as the value
of the previous spins. Thus, for each $2^L$ weights $W^1(i,j-1)$, we
have to store the configuration 
\be
{\cal C}_{S(i,j-1),\cdots,S(i+1,j-2)}
\equiv (S(1,1),\cdots,S(1,L),S(2,1),\cdots,S(i,j-2))\; .
\ee
At the next iteration, we will build the weight $W^1(i,j)$ with the
corresponding configuration
\be
{\cal C}_{S(i,j),\cdots,S(i+1,j-1)}=({\cal C}_{S(i,j-1),\cdots,S(i+1,j-2)},
S(i,j-1))
\ee
with $S(i,j-1)$ the value of the spin which corresponds to the maximum
in (\ref{tfm41}). From this construction, one have access to the
spins configurations of the ground state and the first excited state.

Finally, one should also add that this construction can be easily
extended to the second excited state, etc. After (\ref{tfm40}), one
can easily define a third weight which would be associated to the
second excited state, and so on.

{\bf Acknowledgments.}  We are grateful to J.-P. Bouchaud, D. Huse,
O. Martin, M. A. Moore and A. P. Young for useful comments.  We have
been supported by the Spanish Ministerio de Ciencia y Tecnolog\'{\i}a,
projects BFM2001-3525 (F.R.)  and grant AP98-36523875 (M.S.). M. P. and
F. R. acknowledge support from the French-Spanish collaboration (Picasso
program and Acciones Integradas HF1998-0097). Funding from the European
Science Foundation through the SPHINX program is also acknowledged.

\end{document}